# Empirical Mass-Loss Rates across the Upper Hertzsprung-Russell-Diagram

Claus Leitherer

*Space Telescope Science Institute, 3700 San Martin Dr., Baltimore, MD 21218*

**Abstract**. I provide an overview of the empirical mass-loss rates of hot and cool luminous stars. Stellar species included in this talk are luminous OB stars, Wolf-Rayet stars, asymptotic giant branch stars, and red supergiants. I discuss the scaling of mass loss with stellar properties, with special emphasis on the influences of chemical abundances. Observational errors and systematic uncertainties are still substantial and vary with stellar type. These uncertainties are a major impediment for the construction of reliable stellar evolution models.

## 1. Introduction

The importance of stellar winds and the associated loss of stellar mass had been recognized as early as 1929 when Beals (1929) interpreted the broad Wolf-Rayet (W-R) emission lines in terms of outflowing gas leaving the underlying star. Obtaining a quantitative measurement, or even a rough estimate of the rate of mass loss was – and still is – a challenge. Coming up with a quick ball-park number for the typical stellar mass-loss rate *of an average star in the universe* is in fact rather simple: if the Hubble Deep Field (Williams et al. 1996) or any other related deep survey is used for guidance on the typical galaxy morphology and density in the universe across time and space (which may not be true because of cosmic variance), one can estimate a total of $\sim 10^{23}$ stars currently existing in the universe. Their masses are skewed towards low-mass, solar-type stars. Stellar evolution theory tells us that solar-type stars evolve into white dwarfs with masses of $\sim 0.7$ $M_\odot$. This then suggests a total mass loss of $\sim 0.3$ $M_\odot$ over the stellar lifetime of 10 Gyr. Therefore the typical mass-loss rate of a typical star currently populating the universe is a few times $10^{-11}$ $M_\odot$ yr$^{-1}$, and the total rate for all stars is a few times $10^{12}$ $M_\odot$ yr$^{-1}$.

While this numerical exercise may be instructive, the result is not too useful. Even worse, it is misleading because it hides the important contribution of massive cool and hot stars, which exist in small numbers but whose mass-loss rates ($\dot{M}$) can sometimes be orders of magnitude higher than the above value. $\dot{M}$ of stars with masses higher than approximately 5 $M_\odot$ are the focus of this review. Such stars populate the upper Hertzsprung-Russell diagram (HRD) above $\sim 10^4$ $L_\odot$ and have $\dot{M} \approx 10^{-7} - 10^{-5}$ $M_\odot$ yr$^{-1}$. It is those stars that are primary agents of the galactic interstellar medium, both via their release of matter and kinetic energy. The goal of this review is an assessment of our current knowledge of the stellar mass-loss rates in non-eruptive phases. Outbursts associated with supernovae and Luminous Blue Variables (LBV) are discussed elsewhere.

## 2. Empirical mass-loss rates of hot stars

Massive, hot stars develop powerful stellar winds driven by radiation pressure. Copious ultraviolet (UV) photons emitted in the photospheres are absorbed by metal lines in the outer atmospheric layers, and the associated transfer of momentum from the photons to the ions supports outflows with velocities of order $10^3$ km s$^{-1}$. The original quantitative framework of the radiatively driven wind theory was formulated in a seminal paper by Castor et al. (1975) and has been refined primarily by the Boulder and Munich groups (Kudritzki & Puls 2000).

Earlier observations seemed to confirm the mass-loss rates predicted by the theory, which was then regarded as a general validation. Consequently, the observed O-star rates with values $\dot{M} \approx 10^{-6} - 10^{-5}$ M$_\odot$ yr$^{-1}$ were adopted in the previous, still widely used release of the Geneva stellar evolution models (Meynet et al. 1994).

O-star mass-loss rate determinations fall into three major categories: Hα (or other recombination lines), thermal free-free emission at mid-infrared (IR) or radio wavelengths, and UV resonance lines. Each of the three techniques has its distinct pros and cons but it is the comparison of the UV with the two other methods that reveals the fundamental challenge that is faced by mass-loss determinations. Most Hα and free-free techniques result in $\dot{M}$ values that are consistent with each other (Lamers & Leitherer 1993; Leitherer et al. 1995; Scuderi et al. 1998). This suggests that complications like the a priori unknown run of velocity with photospheric distance, contamination by other emission lines or by non-thermal radiation, and data quality are not an issue. An important result of these studies is the strong luminosity dependence of $\dot{M}$. Since UV photons drive the wind and the same time account for most of the bolometric luminosity ($L$), $\dot{M}$ and $L$ scale as $\dot{M} \propto L^{1.5...2}$.

When looked at in more detail, case studies such as the one of ζ Pup by Puls et al. (2006) suggest a fundamental limitation of techniques which rely on the wind emission measure: they are subject to the wind filling factor. Radiatively driven winds are intrinsically unstable and develop high-contrast clumps and density variations (Owocki 2003). Observational evidence for wind clumping was originally presented by Moffat & Robert (1994). Puls et al. determined that the wind density of ζ Pup is overestimated by a factor of 10 if clumping is neglected, and the corresponding mass-loss rate is too high by a factor of ~3.

A UV study by Fullerton et al. (2006) sheds more light on the effects of clumping on the derived $\dot{M}$. The UV technique utilizes absorption lines, which have linear density dependence, whereas methods relying on the wind emission (such as Hα and the radio) scale with density squared. Therefore a comparison of $\dot{M}$ derived from the two methods will reveal the clumping via the different density dependence. This test is done best with the P V doublet at 1118 and 1128 Å because the line is unsaturated, P$^{4+}$ is the dominant ionization stage, and phosphorus is not subject to nucleosynthetic processing. The outcome is in Fig. 1. The comparison between the emission- and absorption based mass-loss rates suggests a difference of a factor of ~100, which translates into a downward revision of $\dot{M}$ by a factor of 10.

W-R stars are the evolved descendents of massive O stars, and their winds largely mirror the properties of O-star winds. It is now generally accepted that radiation pressure is the driving agent of W-R winds as well. The absence of absorption lines in the spectra of W-R stars precludes UV resonance based techniques for determining $\dot{M}$. Clumping is thought to affect the derived rates in a fashion similar to that in O stars. Observational evidence comes from the comparison of the rates derived from emission lines and from the broad electron scattering wings. The latter have a linear density dependence, and a test analogous to the absorption line method can be devised (Hillier 1991).

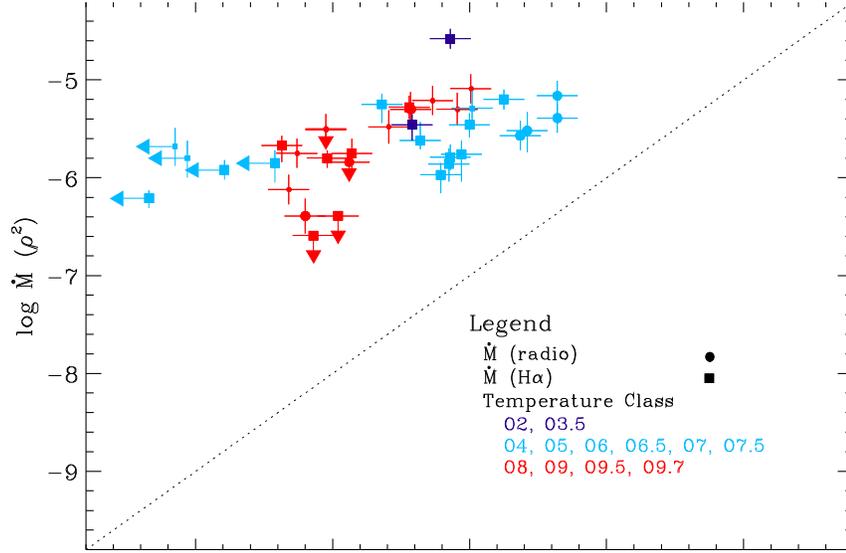

Figure 1. Comparison of density-squared sensitive $\dot{M}$ with $\dot{M}$ from the P V line. The shapes of symbols distinguish radio (circles) and Hα (squares) measurements. Upper limits on non-detections are indicated by arrows. The sample is divided into early (O2, O3, O3.5), mid (O4–O7.5), and late O types (O8–O9.7). From Fullerton et al. (2006).

Crowther (2007) summarized the currently preferred mass-loss rates of W-R stars for each sub-type. The average rates of $\dot{M} \approx 1$–$2 \times 10^{-5}$ M$_\odot$ yr$^{-1}$ exhibit little dependence on spectral type. These values are about a factor of 3 lower than in the unclumped case, similar to the situation for O stars.

The downward revision of the mass-loss rates of unevolved and evolved luminous, massive stars poses a severe challenge for the standard stellar evolution paradigm, which links O main-sequence and evolved W-R stars via strong mass loss (Conti 1976). Consider a 40 M$_\odot$ O star on the main-sequence which is thought to become a WC star on a time-scale of 5 Myr. Assuming a typical WC mass of 15 M$_\odot$, the average $\dot{M}$ prior to entering the W-R phase should be $\sim 10^{-5}$ M$_\odot$ yr$^{-1}$, almost an order of magnitude higher than the clumping corrected O-star rates.

Smith (2008) emphasized the potential role of LBV winds and eruptions. During this short evolutionary phase massive stars could lose as much mass as during the entire O-star phase with a steady wind. Smith's scenario is shown schematically in Fig. 2. In the past, LBVs had been considered has an additional mass-loss sink in stellar evolution. After the downward revision of the O-star mass-loss rates LBVs may turn out to be the single most important mechanism for stellar mass loss prior to the W-R phase. I will return to this issue in Section 5.

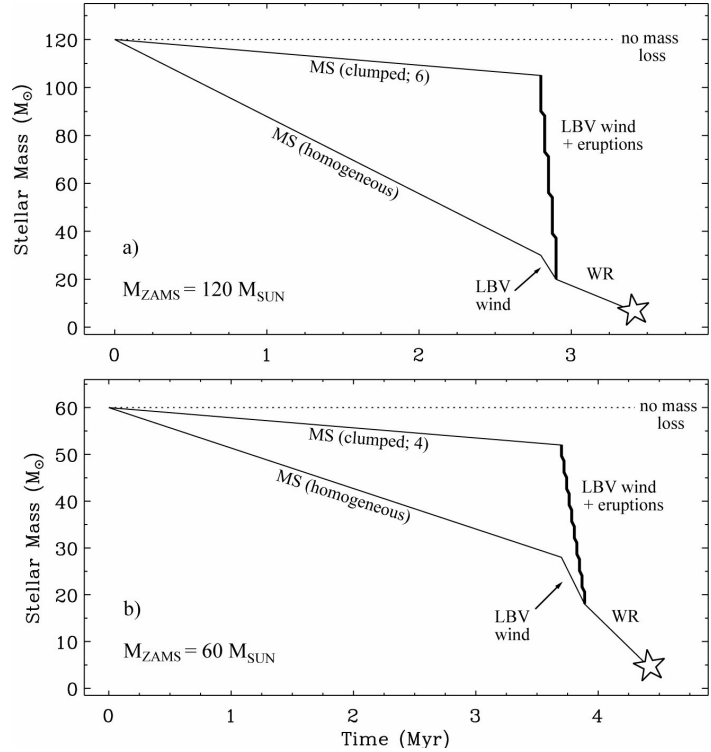

Figure 2. Schematic representation of a star's mass as a function of time. Two extreme scenarios are shown: One has higher conventional O-star mass loss rates assuming homogeneous winds on the main-sequence with no clumping. This is followed by a brief LBV wind phase and a longer W-R wind phase. The second has much reduced $\dot{M}$ on the main sequence (assuming clumping factors of 4–6), followed by an LBV phase that includes severe mass loss in brief eruptions plus a steady wind. Top: 120 $M_\odot$; bottom: 60 $M_\odot$. From Smith (2008).

## 3. Empirical mass-loss rates of cool stars

The upper right part of the HRD is populated by cool, luminous red supergiants (RSG) and symptotic giant branch (AGB) stars. Although RSGs and AGB stars form via different evolutionary channels, their positions in the HRD are partially overlapping between $4.5 < L/L_\odot < 5.0$. The revision of RSG parameters towards lower luminosities discussed at this meeting (see E. Levesque's talk) diminishes the separation between RSGs and AGB stars even further. More importantly, both stellar species share a common wind mechanism. Therefore they will be discussed concurrently.

The winds around cool, luminous stars are driven by stellar pulsation and radiation pressure on dust grains, which results in an extended molecular and dust envelope around the mass losing star (Willson 2007). Shock waves due to pulsation act as a levitation mechanism for the atmosphere, with radiation pressure on dust grains (and possibly on molecules) providing the accelerating force.

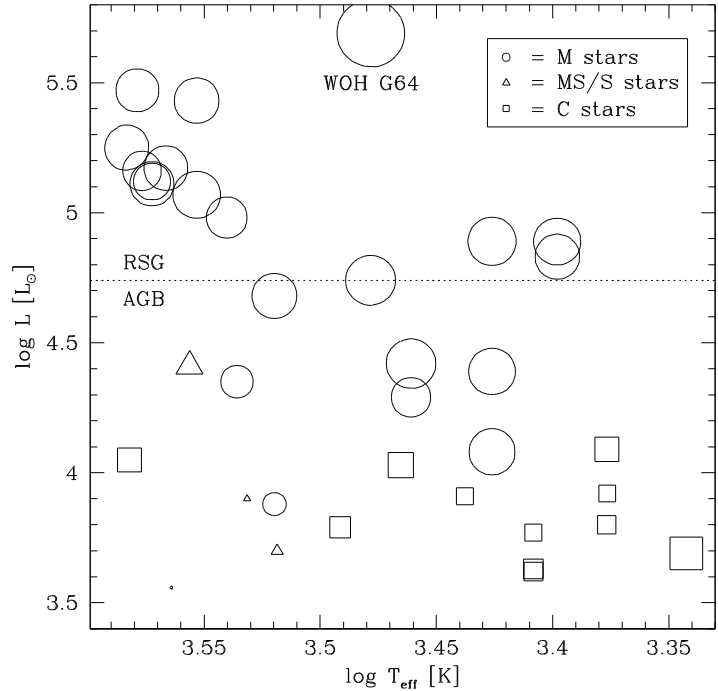

Figure 3. HRD for M-type stars (circles), MS or S-type stars (triangles), and carbon stars (squares), where the sizes of the symbols are logarithmically proportional to the mass-loss rate. From van Loon et al. (2005).

The dust which forms around these cool stars absorbs the visual stellar light and re-emits it in the IR. Even for moderately strong $\dot{M}$ the IR excess can become so prominent that visually faint stars are bright near- and mid-IR sources. The optical depth of the dust can be determined from the IR luminosity and spectral energy distribution and depends on $\dot{M}$, the expansion velocity, $L$, stellar temperature, the dust condensation temperature and the chemistry. The mass-loss rate then follows if the other parameters can be derived independently. This method has been widely used, in particular with large ground- and space-based surveys (e.g., Groenewegen et al. 2007). Typical mass-loss rates for cool stars in the $L \approx 10^4$ $L_\odot$ range are $\sim 10^{-5}$ $M_\odot$ yr$^{-1}$.

van Loon et al. (2005) used this technique to study the scaling of mass loss with basic stellar parameters and to establish an empirical mass-loss relation for RSGs and AGB stars in the Large Magellanic Cloud (LMC). The derived rates are plotted versus stellar effective temperature ($T_{eff}$) in Fig. 3. The strong $T_{eff}$ dependence of $\dot{M}$ is the result of dust destruction at higher temperature and the corresponding diminished driving mechanism. Note that this behavior is opposite to the situation of a hot-star wind, whose $\dot{M}$ has a positive correlation with $T_{eff}$ because higher $T_{eff}$ improves the synchronization of the stellar peak flux wavelength and the location of the driving absorption lines.

Taking into account the $L$ dependence of the mass-loss rate, van Loon et al. (2005) derive for their sample of LMC RSGs and AGB stars the scaling relation

$$\log \dot{M} = -5.65 + 1.05 \log L - 6.3 \log T_{eff},$$

where $\dot{M}$ is in $M_\odot$ yr$^{-1}$, $L$ in $10^4$ $L_\odot$, and $T_{eff}$ in K. This compares fairly well with results for Galactic stars. A major uncertainty that enters in this formula is the assumption of the gas-to-dust ratio of 500. Complementary methods for determining $\dot{M}$ rely, e.g., on the CO emission and are in fair agreement.

Spitzer's unique capabilities allow studies of vastly larger samples. SAGE is a uniform and unbiased imaging survey of the LMC, using the IRAC (3.6, 4.5, 5.8, and 8 μm) and MIPS (24, 70, and 160 μm) instruments. It aims to quantify the rate at which all evolved stars inject mass into the interstellar medium of the LMC (Meixner et al. (2006). Srinivasan et al. (2009) identified about 40,000 AGB star candidates and estimated their mid-IR excesses by comparing them to model photospheres. The excesses serve as a proxy for $\dot{M}$ and reveal general trends of derived parameters such as color temperature and optical depth. Using a grid of dust-shell models, Srinivasan et al. obtained mass-loss rates for the entire sample. The total injection rate of material into the interstellar medium of the LMC from AGB stars is found to be ~1 1 $10^{-4}$ $M_\odot$ yr$^{-1}$.

## 4.  Mass loss in different chemical environments

The arrival of a new generation of powerful observing facilities marked the dawn of the era of "extragalactic stellar astrophysics", where stellar-wind properties can by studied outside our Galaxy and even beyond the Local Group of galaxies. Since the mass-loss rates are predicted to be sensitive to chemical composition, the wind properties should differ from galaxy to galaxy.

Theory predicts a metallicity dependence of $\dot{M} \propto Z^{0.7}$ for hot star winds, where $Z$ denotes the mean abundance of all heavy elements. The most important elements contributing to the line force are C, Si, S, and Fe, with Fe group elements decreasing in importance for higher temperatures. The prediction is hard to test in our Galaxy because of the almost flat metallicity gradient along the disk and the distance uncertainties. The Large (LMC) and Small Magellanic Clouds (SMC) are the obvious choices for this test. One would expect $\dot{M}$ to be lower by about a factor of 1.5 to 2.5 when comparing two otherwise identical O stars in our Galaxy and the LMC or SMC, respectively. Mokiem et al. (2007) tackled this issue both from a theoretical and observational perspective. The comparison is best done using the wind-momentum vs. luminosity relation. This relation reflects the fact that radiative momentum $L/c$ is transferred to kinetic momentum $\dot{M}v$ at an efficiency of ~30% with little variation by stellar type. Mokiem et al. defined a modified wind momentum relation ($D$) which takes into account a weak stellar radius dependence. However, the main parameter affecting $D$ is $\dot{M}$. The key results are reproduced in Fig. 4. The theoretical relation (originally derived by Vink et al. 2001) displays noticeable offsets between the three chosen $Z$ values. The upper panel compares predictions to observed rates which are not corrected for clumping. While the trend of $D$ with $L$ agrees, the observed rates are clearly higher. Since the derived mass-loss rates are most likely overestimated due to the neglect of clumping, a downward correction must be applied. This has been done in the lower panel of Fig. 4. Mokiem et al. chose a correction factor of 2.5 (which must be viewed as quite uncertain). After the correction, theory and observations are in reasonable agreement, at least at higher luminosities.

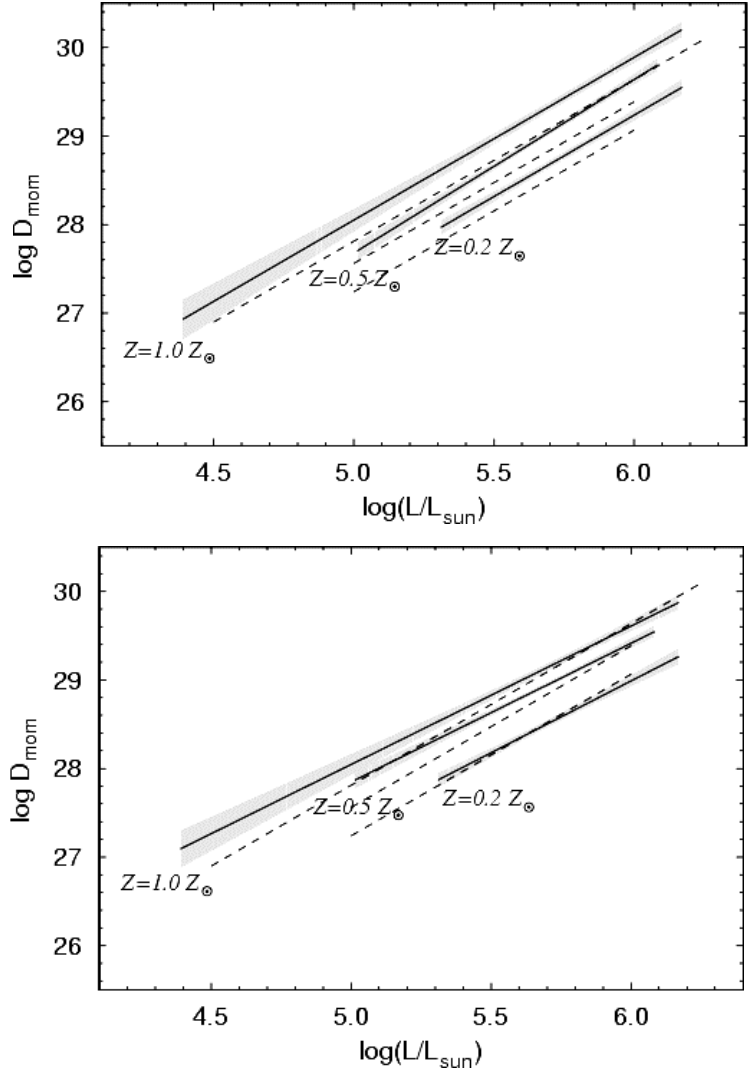

Figure 4. Comparison of the observed wind momentum – luminosity relations (solid lines and shaded areas) with the predicted relations (dashed lines). Top, middle and bottom lines of each line style, respectively, correspond to the Galactic, LMC and SMC observed and predicted relations. The upper and lower panels show uncorrected and clumping corrected observations, respectively. From Mokiem et al. (2007).

The trend of $\dot{M}$ with $Z$ found for O stars appears to apply to W-R stars as well (Crowther 2007). W-R stars in the Galaxy, the LMC, and the SMC form a sequence of decreasing emission line strengths, which can be interpreted as a decrease of $\dot{M}$. The challenge is to disentangle the effects of $Z$ on $\dot{M}$ and on spectral type, which is based on the same emission lines that are used to infer the mass-loss rates.

The major short-coming of studying the $Z$-dependence of $\dot{M}$ in the Magellanic Clouds is the relatively small leverage afforded by the differences in chemical com-position. The oxygen abundance of even the SMC is lower than the solar value by only a factor of 5, which translates into a mass-loss reduction by a factor of three. The only star-forming galaxies in the Local Group with even lower abundance are GR 8, Sextans A, and Leo A, whose oxygen abundances are lower by up to a factor of 4 than the SMC value.

Truly metal-poor star-forming galaxies are more distant and are currently only marginally suitable for extragalactic stellar astrophysics. Two examples of note are DDO 68 and I Zw 18 whose oxygen abundances are ~1/30 the solar value. Izotov & Thuan (2009) reported the discovery of spectral signatures of one or more LBVs in the H II region-like spectrum of DDO 68. A similar discovery was made in the less extreme H II galaxy PHL 293B. Never before have LBVs been detected in such metal-deficient environments. If our standard paradigm of O stars evolving into LBVs by mass loss is correct, the presence of LBVs suggests that mass loss is still not negligible at these very low abundances. A related case can be made with I Zw 18, whose integrated spectrum exhibits signatures of W-R stars. Brown et al. (2002) used HST's STIS to obtain spectra of several individual W-R stars. The tell-tale WC signatures are clearly detected in the spectra. It is currently unknown whether any of these W-R stars are members of close binary systems, which could affect their evolution. If they are not, significant mass loss would have to be invoked for the formation of the WC stars, which have generally low mass.

## 5. Mass-loss rates in stellar evolution models

If our basic understanding of the evolutionary connections between different stellar species in the upper HRD is correct, strong mass loss must occur in order to decrease the mass of an initially massive O star to that of a much less massive W-R star or RSG. Until recently, observational mass-loss rates were thought to be sufficiently well understood that they could serve as a cornerstone for evolution models (Maeder & Conti 1994). The discovery of wind inhomogeneities and the importance of eruptive phase both on the blue and the red side of the HRD have dramatically changed this perspective. The uncertainties of the observed rates can easily reach factors of several. Therefore $\dot{M}$ in evolution models is now considered an adjustable parameter that together with rotation governs the stellar properties with time.

The mass-loss rates adopted in the latest set of Geneva evolution models with rotation are reproduced in Fig. 5 (Meynet & Maeder 2003). The two stellar models are for a 60 $M_\odot$ star, which corresponds to an early-O star on the main-sequence, and a 20 $M_\odot$ star, which would appear as a late-O star. The mass-loss history in the O-star phase is described by the theoretical predictions of Vink et al (2001), modified by rotational mass loss. The 60 $M_\odot$ star subsequently enters the LBV phase with a short outburst (for zero rotation) and then evolves into a W-R star. These models assume particular strong mass loss in the W-R stage to compensate for the low $\dot{M}$ early on the main-sequence. The W-R rates follow from the empirical relation of Nugis & Lamers (2000) who applied a clumping correction. Consistent with observations, less than 25% of the stellar mass is lost during the O star phase. The 20 $M_\odot$ star does not go through the LBV and W-R phase but evolves to the red side of the HRD to become an RSG. As in the prior example, most of the mass loss occurs at very late phases, in this case as an RSG. The RSG mass-loss rates adopted in these particular evolution models are those of de Jager et al. (1988).

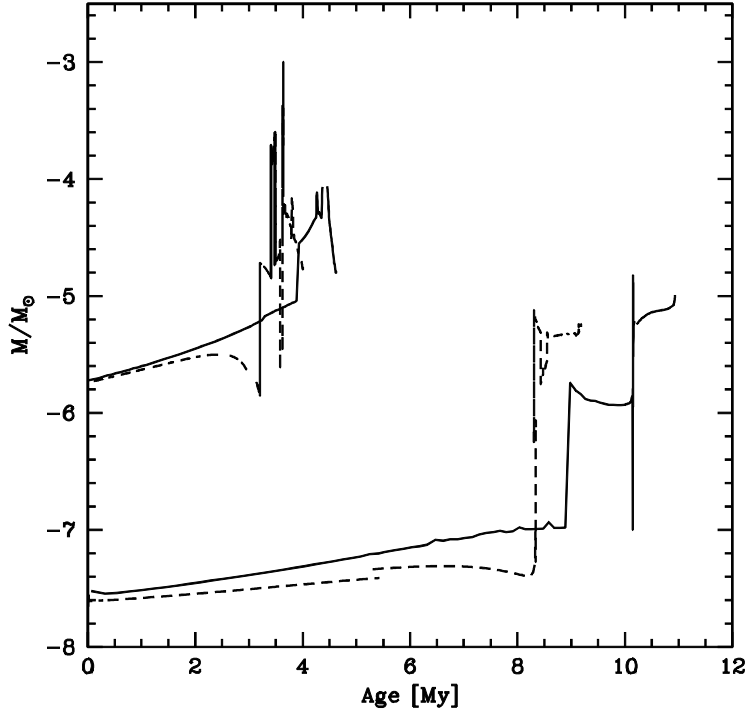

Figure 5. Mass-loss rates used in the latest set of Geneva models versus stellar age. The two sets of graphs are for stars with initial mass 60 (upper) and 20 $M_\odot$ (lower), both rotating with 300 km s$^{-1}$ on the zero-age main-sequence. Models with rotation are solid lines. For comparison, non-rotating models are plotted with dashed lines. Solar chemical composition (Meynet, private communication).

The mass-loss rates adopted in these models are overall consistent with empirical data but the uncertainties are large. Both the precision and the accuracy of observed $\dot{M}$ values for hot and cool luminous stars lack the level that is needed to provide reliable guidance for stellar evolution models. Vice versa, matching stellar population properties predicted by evolution models and observations can help to constrain the effect of mass loss on evolutionary tracks.

**Acknowledgements**. Georges Meynet kindly provided Fig. 5 as well as invaluable insight into the Geneva stellar evolution models. I also thank Sundar Srinivasan for making some results of this thesis work available to me prior to publication.

**References**

Beals, C. S. 1929, MNRAS, 90, 202